\documentclass[conference]{IEEEtran}
\usepackage{dsfont,graphicx,color,epsf,epsfig,verbatim,amssymb,amsmath,amsthm}
\usepackage{latexsym,amsfonts,caption,array,cite,subfigure,multicol,multirow}

\newcommand{\argmax}{\operatornamewithlimits{argmax}}

\setkeys{Gin}{draft=false}

\newtheorem{lemma}{Lemma}[section]

\newtheorem{theorem}{Theorem}[section]
\newtheorem{definition}{Definition}[section]

\newcommand{\G}{\mathbf{G}}

\begin{document}
\title{Nested Polar Codes Achieve the Shannon Rate-Distortion Function and the Shannon Capacity}% can use linebreaks \\
\author{\IEEEauthorblockN{Aria G. Sahebi and S. Sandeep Pradhan\\ \thanks{This work was supported by NSF grants CCF-0915619 and CCF-1116021.}}
\IEEEauthorblockA{Department of Electrical Engineering and Computer Science,\\
University of Michigan, Ann Arbor, MI 48109, USA.\\
Email: \tt\small ariaghs@umich.edu, pradhanv@umich.edu}}

%\markboth{Draft}
%{Sahebi \MakeLowercase{\textit{et al.}}: Multilevel Polarization of Polar Codes Over Arbitrary Discrete Memoryless Channels}

\maketitle

\begin{abstract}
It is shown that nested polar codes achieve the Shannon capacity of arbitrary discrete memoryless sources and the Shannon capacity of arbitrary discrete memory less channels.
\end{abstract}

\section{Introduction}
Polar codes were originally proposed by Arikan in \cite{arikan_polar} to achieve the symmetric capacity of binary-input discrete memoryless channels. Polar codes for lossy source coding were investigated in \cite{korada_lossy_source} where it is shown that polar codes achieve the symmetric rate-distortion function for sources with binary reconstruction alphabets. For the lossless source coding problem, the source polarization phenomenon is introduced in \cite{arikan_source_polarization} to compress a source down to its entropy.

It is well known that linear codes can at most achieve the symmetric capacity of discrete memoryless channels and the symmetric rate-distortion function for discrete memoryless sources. This indicates that polar codes are optimal linear codes in terms of the achievable rate. It is also known that nested linear codes achieve the Shannon capacity of arbitrary discrete memoryless channels and the Shannon rate-distortion function for arbitrary discrete memoryless sources. In this paper, we investigate the performance of nested polar codes for the point-to-point channel and source coding problems and show that these codes achieve the Shannon capacity of arbitrary (binary or non-binary) DMCs and the Shannon rate-distortion function for arbitrary DMSs.

The results of this paper are general regarding the size of the channel and source alphabets. To generalize the results to non-binary cases, we use the approach of \cite{sahebi_multilevel_polar_ieee} in which it is shown that polar codes with their original $(u,u+v)$ kernel, achieve the symmetric capacity of arbitrary discrete memoryless channels where $+$ is the addition operation over any finite Abelian group.
\section{Preliminaries} \label{prel}
\subsubsection{Source and Channel Models}
For the source coding problem, the source is modeled as a discrete-time random process with each sample taking values in a fixed finite set $\mathcal{X}$ with probability distribution $p_X$. The reconstruction alphabet is denoted by $\mathcal{U}$ and the quality of reconstruction is measured by a single-letter distortion function $d:\mathcal{X}\times \mathcal{U}\rightarrow \mathds{R}^{+}$. We denote the source by $(\mathcal{X},\mathcal{U},p_{X},d)$. With a slight abuse of notation, for $x^n\in\mathcal{X}^n$ and $u^n\in\mathcal{U}^n$, we define
\begin{align*}
d(x^n,u^n)=\frac{1}{n}\sum_{i=1}^n d(x_i,u_i)
\end{align*}
For the channel coding problem, we consider discrete memoryless and stationary channels used without feedback. We associate two finite sets $\mathcal{X}$ and $\mathcal{Y}$ with the channel as the channel input and output alphabets. These channels can be characterized by a conditional probability law $W(y|x)$ for $x\in \mathcal{X}$ and $y\in \mathcal{Y}$. The channel is specified by $(\mathcal{X},\mathcal{Y},W)$. The source of information generates messages over the set $\{1,2,\ldots,M\}$ uniformly for some positive integer $M$.\\
\subsubsection{Achievability and the Rate-Distortion Function for the Source Coding Problem}
A transmission system with parameters $(n,\Theta,\Delta,\tau)$ for compressing a given source $(\mathcal{X},\mathcal{U},p_{X},d)$ consists of an encoding mapping and a decoding mapping %$e:\mathcal{X}^n\rightarrow \{1,2,\cdots,\Theta\}$ and a decoding mapping $g:\mathcal{S}^n\times \{1,2,\cdots,\Theta\}\rightarrow \mathcal{U}^n$
\begin{align*}
&\mbox{\small Enc}:\mathcal{X}^n\rightarrow \{1,2,\cdots,\Theta\},\\
&\mbox{\small Dec}:\{1,2,\cdots,\Theta\}\rightarrow \mathcal{U}^n
\end{align*}
such that the following condition is met:%$P\left(d(X^n,g(e(X^n)))>\Delta\right)\le \tau$ %
\begin{align*}
P\left(d\left(X^n,\mbox{\small Dec}(\mbox{\small Enc}(X^n))\right)>\Delta\right)\le \tau
\end{align*}
where $X^n$ is the random vector of length $n$ generated by the source. In this transmission system, $n$ denotes the block length, $\log \Theta$ denotes the number of channel uses, $\Delta$ denotes the distortion level and $\tau$ denotes the probability of exceeding the distortion level $\Delta$.\\
Given a source, a pair of non-negative real numbers $(R,D)$ is said to be achievable if there exists for every $\epsilon>0$, and for all sufficiently large numbers $n$ a transmission system with parameters $(n,\Theta,\Delta,\tau)$ for compressing the source such that
\begin{align*}
\frac{1}{n}\log \Theta\le R+\epsilon, \qquad \Delta\le D+\epsilon,\qquad \tau\le \epsilon
\end{align*}
The optimal rate distortion function $R^*(D)$ of the source is given by the infimum of the rates $R$ such that $(R,D)$ is achievable.\\
It is known that the optimal rate-distortion function is given by:
\begin{align}\label{eqn:Shannon_RD}
R(D)=\inf_{\substack{p_{U|X}\\\mathds{E}_{p_X p_{U|X}}\{d(X,U)\}\le D}} I(X;U)
\end{align}
where $p_{U|X}$ is the conditional probability of $U$ given $X$. % and $p_X p_{U|X}$ is the joint distribution of $X$ and $U$.\\
%The \emph{symmetric rate-distortion} function $\bar{R}(D)$ is defined as follows:
%\begin{align*}
%\bar{R}(D)=\min_{\substack{p_{U|X}\\\mathds{E}_{p_X p_{U|X}}\{d(X,U)\}\le D\\p_U=\frac{1}{q}}} I(X;U)
%\end{align*}
%where $p_U$ is the marginal distribution of $U$ given by $p_U(u)=\sum_{x\in\mathcal{X}}p_X(x) p_{U|X}(u|x)$ and $q$ is the size of the reconstruction alphabet $\mathcal{U}$.

\subsubsection{Achievability and Capacity for the Channel Coding Problem}
A transmission system with parameters $(n,M,\tau)$ for reliable communication over a given channel $(\mathcal{X},\mathcal{Y},W)$ consists of an encoding mapping $\mbox{\small Enc}:\{1,2,\ldots,M\}\rightarrow \mathcal{X}^n$ and a decoding mapping $\mbox{\small Dec}:\mathcal{Y}^n\rightarrow\{1,2,\ldots,M\}$ such that
\begin{align*}
\frac{1}{M}\sum_{m=1}^{M} W^n\left(\mbox{\small Dec}(Y^n)\ne
m|X^n=\mbox{\small Enc}(m)\right)\le \tau
\end{align*}
Given a channel $(\mathcal{X},\mathcal{Y},W)$, the rate $R$ is said to be achievable if for all $\epsilon>0$ and for all sufficiently large $n$, there exists a transmission system for reliable communication with parameters $(n,M,\tau)$ such that
\begin{align*}
\frac{1}{n}\log M \ge R-\epsilon,\qquad\qquad \tau\le \epsilon
\end{align*}
The channel capacity is the supremum of the set of achievable rates. It is known that the channel capacity is given by:
\begin{align}\label{eqn:Shannon_C}
C=\sup_{p_{X}} I(X;Y)
\end{align}
where $p_{X}$ is the channel input distribution.\\

\subsubsection{Groups, Rings and Fields}
All groups referred to in this paper are \emph{Abelian groups}. Given a group $(\G,+)$, a subset $H$ of $\G$ is called a \emph{subgroup} of $\G$ if it is closed under the group operation. In this case, $(H,+)$ is a group in its own right. This is denoted by $H\le \G$. A \emph{coset} $C$ of a subgroup $H$ is a shift of $H$ by an arbitrary element $a\in \G$ (i.e. $C=a+H$ for some $a\in\G$). For any subgroup $H$ of $\G$, its cosets partition the group $\G$. A \emph{transversal} $T$ of a subgroup $H$ of $\G$ is a subset of $\G$ containing one and only one element from each coset (shift) of $H$. Given an element $d$ of $\G$, $\langle d\rangle$ denotes the subgroup of $\G$ generated by $d$. i.e. the smallest subgroup of $\G$ containing $d$. A subgroup $M$ of $\G$ is called maximal if it is a proper subgroup and there does not exist another proper subgroup of $\G$ containing $M$.\\
%We give some examples in the following: The simplest non-trivial example of groups is $\mathds{Z}_2$ with addition mod-$2$ which is a \emph{ring} and a \emph{field} with multiplication mod-$2$. The group $\mathds{Z}_2\times \mathds{Z}_2$ is also a ring and a field under component-wise mod-$2$ addition and a carefully defined multiplication. The group $\mathds{Z}_4$ with mod-$4$ addition and multiplication is a ring with prime-power order but not a field since the element $2\in\mathds{Z}_4$ does not have a multiplicative inverse. The subset $\{0,2\}$ is a subgroup of $\mathds{Z}_4$ since it is closed under mod-$4$ addition. $\{0\}$ and $\mathds{Z}_4$ are the two other subgroups of $\mathds{Z}_4$. The group $\mathds{Z}_6$ is neither a field nor a ring with prime-power order. Subgroups of $\mathds{Z}_6$ are: $\{0\}$, $\{0,3\}$, $\{0,2,4\}$ and $\mathds{Z}_6$.\\

\subsubsection{Channel Parameters}
For a channel $(\mathcal{X},\mathcal{Y},W)$, assume $\mathcal{X}$ is equipped with the structure of a group $(\G,+)$. The symmetric capacity is defined as $\bar{I}(W)=I(X;Y)$ where the channel input $X$ is uniformly distributed over $\mathcal{X}$ and $Y$ is the output of the channel. %; i.e. for $q=|\mathcal{X}|$,
%\begin{align*}
%I^0(W)=\sum_{x\in\mathcal{X}}\sum_{y\in\mathcal{Y}}\frac{1}{q}W(y|x)\log\frac{W(y|x)}{\displaystyle\sum_{{\tilde{x} \in\mathcal{X}}}\frac{1}{q}W(y|\tilde{x})}
%\end{align*}
The Bhattacharyya distance between two distinct input symbols $x$ and $\tilde{x}$ is defined as
\begin{align*}
Z(W_{\{x,\tilde{x}\}})=\sum_{y\in\mathcal{Y}}\sqrt{W(y|x)W(y|\tilde{x})}
\end{align*}
and the average Bhattacharyya distance is defined as
\begin{align*}
Z(W)=\sum_{\substack{x,\tilde{x}\in \mathcal{X}\\x\ne\tilde{x}}}\frac{1}{q(q-1)}Z(W_{\{x,\tilde{x}\}})
\end{align*}
where $q=|\mathcal{X}|$. We use the following two quantities in the paper extensively:
\begin{align*}
&D_d(W)=\frac{1}{2q}\sum_{u\in \mathcal{U}} \sum_{x\in\mathcal{X}} \left|W(x|u)-W(x|u+d)\right|\\
&\tilde{D}_d(W)=\frac{1}{2q}\sum_{u\in \mathcal{U}} \sum_{x\in\mathcal{X}} \left(W(x|u)-W(x|u+d)\right)^2
\end{align*}
where $d$ is some element of $\G$ and $+$ is the group operation.\\

\subsubsection{Binary Polar Codes}
For any $N=2^n$, a polar code of length $N$ designed for the channel $(\mathds{Z}_2,\mathcal{Y},W)$ is a linear (coset) code characterized by a generator matrix $G_N$ and a set of indices $A\subseteq \{1,\cdots,N\}$ of \emph{almost perfect channels}. The generator matrix for polar codes is defined as $G_N=B_NF^{\otimes n}$ where $B_N$ is a permutation of rows, $F=\left[ \begin{array}{cc}1 & 0\\1 & 1\end{array} \right]$ and $\otimes$ denotes the Kronecker product. The set $A$ is a function of the channel. The decoding algorithm for polar codes is a specific form of successive cancellation \cite{arikan_polar}.\\

\subsubsection{Polar Codes Over Abelian Groups}
For any discrete memoryless channel, there always exists an {Abelian group} of the same size as that of the channel input alphabet. In general, for an Abelian group, there may not exist a multiplication operation. Since polar encoders are characterized by a matrix multiplication, before using these codes for channels of arbitrary input alphabet sizes, a generator matrix for codes over Abelian groups needs to be properly defined. Polar codes over Abelian groups are introduced in \cite{sahebi_multilevel_polar_ieee}.\\

%\subsubsection{Group Codes}
%Let the channel input alphabet $\mathcal{X}$ be equipped with the structure of a finite Abelian group $\G$ of the same size. Then the channel is specified by $(\G,\mathcal{Y},W)$. A group code over $\G$ of length $N$ for this channel is any {subgroup} of $\G^N$. The group capacity of a channel $(\G,\mathcal{Y},W)$ is the maximum achievable rate using group codes over $\G$ for this channel. Group codes generalize the notion of linear codes over {fields} to channels with composite input alphabet sizes. A coset code is a shift of a group code by a constant vector.\\

\subsubsection{Notation}%We say a function $f:\mathds{R}\rightarrow\mathds{R}$ is $O(\epsilon)$ if it's right limit is zero at zero.
We denote by $O(\epsilon)$ any function of $\epsilon$ which is right-continuous around $0$ and that $O(\epsilon)\rightarrow 0$ as $\epsilon\downarrow 0$.\\
For positive integers $N$ and $r$, let $\{A_0,A_1,\cdots,A_r\}$ be a partition of the index set $\{1,2,\cdots,N\}$. Given sets $T_t$ for $t=0,\cdots,r$, the direct sum $\bigoplus_{t=0}^r T_t^{A_t}$ is defined as the set of all tuples $u_1^N=(u_1,\cdots,u_N)$ such that $u_i\in T_t$ whenever $i\in A_t$.\\

\section{The Lossy Source Coding Problem}
In this section, we prove the following theorem:
\begin{theorem}
For an arbitrary discrete memoryless source $(\mathcal{X},\mathcal{U},p_X,d)$, nested polar codes achieve the Shannon rate-distortion function \eqref{eqn:Shannon_C}.
\end{theorem}

For the source $(\mathcal{X},\mathcal{U},p_X,d)$, let $\mathcal{U}=\G$ where $\G$ is an arbitrary Abelian group and let $q=|\G|$ be the size of the group. For a pair $(R,D)\in\mathds{R}^2$, let $X$ be distributed according to $p_X$ and let $U$ be a random variable such that $\mathds{E}\{d(X,U)\}\le D$. We prove that there exists a pair of polar codes $\mathds{C}_i\subseteq\mathds{C}_o$ such that $\mathds{C}_i$ induces a partition of $\mathds{C}_o$ through its shifts, $\mathds{C}_o$ is a good source code for $X$ and each shift of $\mathds{C}_i$ is a good channel code for the test channel $p_{X|U}$. This will be made clear later in the following.\\

\begin{figure}[!h]
\centering
\includegraphics[scale=1]{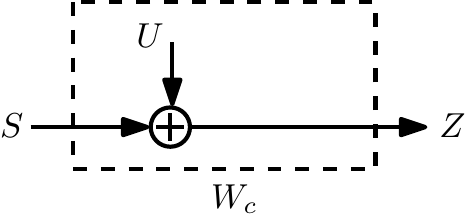}
\caption{\small Test channel for the inner code (the channel coding component)}
\label{fig:Wc_RD}
\end{figure}

\begin{figure}[!h]
\centering
\includegraphics[scale=1]{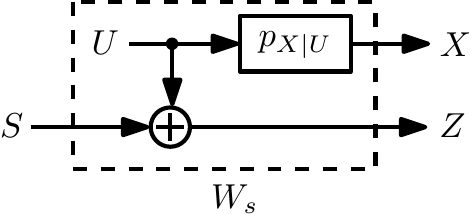}
\caption{\small Test channel for the outer code (the source coding component)}
\label{fig:Ws_RD}
\end{figure}

Given the test channel $p_{X|U}$, define the artificial channels $(\G,\G,W_c)$ and $(\G,\mathcal{X}\times\G,W_s)$ such that for $s,z\in \G$ and $x\in \mathcal{X}$,
\begin{align*}
&W_c(z|s)=p_U(z-s)\\
&W_s(x,z|s)=p_{XU}(x,z-s)
\end{align*}
These channels have been depicted in Figures \ref{fig:Wc_RD} and \ref{fig:Ws_RD}.

Let $S$ be a random variable uniformly distributed over $\G$ which is independent from $X$ and $U$. It is straightforward to show that in this case, $Z$ is also uniformly distributed over $\G$. The symmetric capacity of the channel $W_c$ is equal to
\begin{align*}
\bar{I}(W_c)=I(S;Z)&=H(Z)-H(Z|S)\\
&\stackrel{(a)}{=}\log q-H(U|S)=\log q-H(U)
\end{align*}
where $(a)$ follows since $Z$ is uniformly distributed and $U$ is independent of $S$. For the channel $W_s$, first we show that $X$ and $Z$ are independent. For $z\in \G$ and $x\in\mathcal{X}$,
\begin{align*}
p_{X|Z}(x|z)&=\sum_{u\in \G} p_{U|Z}(u|z) p_{X|ZU}(x|z,u)\\
&=\sum_{u\in \G} \frac{p_{US}(u,z-u)}{p_{Z}(z)} p_{X|ZU}(x|z,u)\\
&\stackrel{(a)}{=}\sum_{u\in \G}  p_U(u) p_{X|U}(x|u)\\
&=p_X(x)
\end{align*}
where $(a)$ follows since $S$ and $U$ are independent, $S$ and $Z$ are uniformly distributed and the Markov chain $Z\leftrightarrow U \leftrightarrow X$ holds. The symmetric capacity of the channel $W_s$ is equal to
\begin{align*}
\bar{I}(W_s)=I(S;XZ)&=H(S)+H(XZ)-H(SXZ)\\
&\stackrel{(a)}{=}H(S)+H(X)+H(Z)-H(SXU)\\
&\stackrel{(b)}{=}H(X)+H(Z)-H(XU)\\
&\stackrel{(c)}{=}\log q-H(U|X)\\
\end{align*}
where $(a)$ follows since $X$ and $Z$ are independent and there is a one-to-one correspondence between $(S,Z)$ and $(S,U)$. Equality $(b)$ follows since $S$ is independent of $X,U$ and hence $H(SXU)=H(S)+H(XU)$. Equality $(c)$ follows since $Z$ is uniform.\\
We employ a nested polar code in which the inner code is a good channel code for the channel $W_c$ and the outer code is a good source code for $W_s$. The rate of this code is equal to
\begin{align*}
R&=\bar{I}(W_s)-\bar{I}(W_c)\\
&=\log q-H(U|X)-\left(\log q-H(U)\right)=I(X;U)
\end{align*}
Note that the channels $W_c$ and $W_s$ are chosen so that the difference of their \emph{symmetric} capacities is equal to the \emph{Shannon} mutual information between $U$ and $X$. This enables us to use channel coding polar codes to achieve the symmetric capacity of $W_c$ (as the inner code) and source coding polar codes to achieve the symmetric capacity of the test channel $W_s$ (as the outer code). The exact proof is postponed to Section \ref{section:Binary_proof_RD} where the result is proved for the binary case and Section \ref{section:General_proof_RD} in which the general proof (for arbitrary Abelian groups) is presented.

The next section is devoted to some general definitions and useful lemmas which are used in the proofs.
\subsection{Definitions and Lemmas}
For a channel $(\mathcal{X},\mathcal{Y},W)$, the basic channel transformations associated with polar codes are given by:
\begin{align}
&\label{eqn:channel_transform1} W^-(y_1,y_2|u_1)=\sum_{u_2^\prime\in \G}\frac{1}{q}W(y_1|u_1+u_2^\prime)W(y_2|u_2^\prime)\\
&\label{eqn:channel_transform2} W^+(y_1,y_2,u_1|u_2)=\frac{1}{q}W(y_1|u_1+u_2)W(y_2|u_2)
\end{align}
for $y_1,y_2\in\mathcal{Y}$ and $u_1,u_2\in \G$. We apply these transformations to both channels $(\G,\G,W_c)$ and $(\G,\mathcal{X}\times\G,W_s)$. Repeating these operations $n$ times recursively for $W_c$ and $W_s$, we obtain $N=2^n$ channels $W_{c,N}^{(1)},\cdots,W_{c,N}^{(N)}$ and $W_{s,N}^{(1)},\cdots,W_{s,N}^{(N)}$ respectively. For $i=1,\cdots,N$, these channels are given by:
\begin{align*}
W_{c,N}^{(i)}(z_1^n,v_1^{i-1}|v_i) &=\sum_{v_{i+1}^N\in \G^{N-i}} \frac{1}{q^{N-1}} W_c^N(z_1^N|v_1^NG)\\
&= \sum_{v_{i+1}^N\in \G^{N-i}} \frac{1}{q^{N-1}} p_U^N(z_1^N-v_1^NG)\\
W_{s,N}^{(i)}(x_1^N,z_1^n,v_1^{i-1}|v_i) &=\sum_{v_{i+1}^N\in \G^{N-i}} \frac{1}{q^{N-1}} W_s^N(x_1^N,z_1^N|v_1^NG)\\
&= \sum_{v_{i+1}^N\in \G^{N-i}} \frac{1}{q^{N-1}} p_{XU}^N(x_1^N,z_1^N-v_1^NG)\\
\end{align*}
for $z_1^N,v_1^N\in \G^N$, $x_1^N\in\mathcal{X}^N$ where $G$ is the generator matrix of dimensions $N\times N$ for polar codes.
%Let $J_n$ be a uniform random variable over the set $\{1,2,\cdots,N\}$ and define
%\begin{align*}
%&I^n(W_c)=\bar{I}(W_{c,N}^{(J_n)})\\
%&I^n(W_s)=\bar{I}(W_{s,N}^{(J_n)})\\
%&Z^n(W_c)=Z(W_{c,N}^{(J_n)})\\
%&Z^n(W_s)=Z(W_{s,N}^{(J_n)})
%\end{align*}
For the case of binary input channels, it has been shown in \cite{arikan_polar} that as $N\rightarrow \infty$, these channels polarize in the sense that their Bhattacharyya parameter gets either close to zero (perfect channels) or close to one (useless channels). For arbitrary channels, it is shown in \cite{sahebi_multilevel_polar_ieee} that polarization happens in multiple levels so that as $N\rightarrow \infty$ channels get useless, perfect or ``partially perfect''.\\
For an integer $n$, let $J_n$ be a uniform random variable over the set $\{1,2,\cdots,N=2^n\}$ and define the random variable $I^n(W)$ as
\begin{align}\label{eqn:Iprocess}
I^n(W)=I(X;Y)
\end{align}
where $X$ and $Y$ are the input and output of $W_N^{(J_n)}$ respectively and $X$ is uniformly distributed. It has been shown in \cite{sasoglu_polar_q} that the process $I^0,I^1,I^2,\cdots$ is a martingale; hence $\mathds{E}\{I^n\}=I^0$. For an integer $n$, define the random variable $Z_d^n(W)=Z_d(W_N^{(J_n)})$ where for a channel $(\G,\mathcal{Y},W)$,
\begin{align}\label{eqn:Zd}
Z_d(W)=\frac{1}{q}\sum_{x\in \G}\sum_{y\in \mathcal{Y}}\sqrt{W(y|x)W(y|x+d)}
\end{align}
Other than the processes $I^n(W)$ and $Z_d^n(W)$, in the proof of polarization, we need another set of processes $Z^H(W)$ and $I^n_H(W)$ for $H\le \G$ which we define in the following. Define
\begin{align*}
Z^H(W)=\sum_{d\notin H} Z_d(W)
\end{align*}
%For an arbitrary $d\in \G$, let $H=\langle d\rangle$ be the subgroup generated by $d$ i.e. the set of all elements of $\G$ which are divisible by $d$.
Note that any uniform random variable defined over $\G$ can be decomposed into two uniform and independent random variables $[X]_H$ and $[X]_{T_H}$ where $[X]_H$ takes values from $H$ and $[X]_{T_H}$ takes values from the transversal $T$ of $H$ such that $X=[X]_H+[X]_{T_H}$. For an integer $n$, define the random variable $I^n_H(W)$ as
\begin{align}\label{eqn:Iprocess2}
I^n_H(W)=I(X;Y|[X]_{T_H})=I([X]_H;Y|[X]_{T_H})
\end{align}

\begin{definition}\label{def:degraded}
The channel $(\G,\mathcal{Y}_1,W_1)$ is degraded with respect to the channel $(\G,\mathcal{Y}_2,W_2)$ if there exists a channel $(\mathcal{Y}_2,\mathcal{Y}_1,W)$ such that for $x\in\G$ and $y_1 \in \mathcal{Y}_1$,
\begin{align*}
W_1(y_1|x)=\sum_{y_2\in\mathcal{Y_2}} W_2(y_2|x) W(y_1|y_2)
\end{align*}
\end{definition}

\begin{lemma}\label{lemma:Degraded_ZZ}
If the channel $(\G,\mathcal{Y}_1,W_1)$ is degraded with respect to the channel $(\G,\mathcal{Y}_2,W_2)$ then $Z(W_1)\ge Z(W_2)$.
\end{lemma}
\begin{IEEEproof}
Follows from [???].
\end{IEEEproof}

\begin{lemma}\label{lemma:Degraded_WW}
If the channel $(\G,\mathcal{Y}_1,W_1)$ is degraded with respect to the channel $(\G,\mathcal{Y}_2,W_2)$ then $(\G,\mathcal{Y}_1\times \mathcal{Y}_1\times \G,W_1^+)$ is degraded with respect to the channel $(\G,\mathcal{Y}_2\times \mathcal{Y}_2\times \G,W_2^+)$ and $(\G,\mathcal{Y}_1\times \mathcal{Y}_1,W_1^-)$ is degraded with respect to the channel $(\G,\mathcal{Y}_2\times \mathcal{Y}_2,W_2^-)$
\end{lemma}
\begin{IEEEproof}
Follows from [???].
\end{IEEEproof}

\begin{lemma}\label{lemma:Wc_deg_Ws_RD2}
The channel $W_c$ is degraded with respect to the channel $W_s$ in the sense of Definition \ref{def:degraded}.
\end{lemma}
\begin{IEEEproof}
Intuitively, it is clear that $W_c$ is a degraded version of $W_s$. The proof is as follows: Let the channel $(\mathcal{X}\times \G,\G,W)$ be such that for $z,z'\in\G$ and $x\in\mathcal{X}$, $W(z|x,z')=\mathds{1}_{\{z=z'\}}$. Then for $s,z\in\G$,
\begin{align*}
\sum_{\substack{z'\in\G\\x\in\mathcal{X}}} W_s(x,z'|s)\mathds{1}_{\{z=z'\}} &= \sum_{\substack{z'\in\G\\x\in\mathcal{X}}} P_{XU}(x,z'-s)\cdot \mathds{1}_{\{z=z'\}}\\
&=\sum_{x\in\mathcal{X}} P_{XU}(x,z'-s)\\
&= P_{U}(x,z'-s) = W_c(z|s)
\end{align*}
\end{IEEEproof}

Let the random vectors $X_1^N,U_1^N$ be distributed according to $P_{XU}^N$ and let $Z_1^N$ be a random variable uniformly distributed over $\G^N$ which is independent of $X_1^N,U_1^N$. Let $S_1^N=Z_1^N-U_1^N$ and $V_1^N=S_1^N G^{-1}$ (Here, $G^{-1}$ is the inverse of the one-two-one mapping $G:\G^N\rightarrow \G^N$). In other words, the joint distribution of these random vectors is given by
\begin{align*}
&p_{V_1^NS_1^NU_1^NX_1^NZ_1^N}(v_1^N,s_1^N,u_1^N,x_1^N,z_1^N)\\
&\qquad\qquad=\frac{1}{q^N} p_{XU}^N(x_1^N,u_1^N) \mathds{1}_{\{s_1^N=v_1^NG,u_1^n=z_1^N-v_1^NG\}}
\end{align*}
This implies
\begin{align*}
p_{V_1^NX_1^NZ_1^N}(v_1^N,x_1^N,z_1^N)&=\frac{1}{q^N} p_{XU}^N(x_1^N,z_1^N-v_1^NG),\\
p_{V_1^NZ_1^N}(v_1^N,z_1^N)&=\frac{1}{q^N} p_{U}^N(z_1^N-v_1^NG)
\end{align*}

In the next section, we provide the proof for the binary case.
\subsection{Source Coding: Sketch of the Proof for the Binary Case}\label{section:Binary_proof_RD}
The standard result of channel polarization for the binary input channel $W_c$ implies \cite{arikan_polar} that for any $\epsilon>0$ and $0< \beta <\frac{1}{2}$, there exist a large $N=2^n$ and a partition $A_0,A_1$ of $[1,N]$ such that for $t=0,1$ and $i\in A_t$, $\left|\bar{I}(W_{c,N}^{(i)}) -t \right|<\epsilon$ and such that for $i\in A_1$ $Z(W_{c,N}^{(i)})< 2^{-N^{\beta}}$. Moreover, as $\epsilon\rightarrow 0$ (and $N\rightarrow \infty$), $\frac{|A_t|}{N}\rightarrow p_t$ for some $p_0,p_1$ adding up to one with $p_1=\bar{I}(W_c)$.\\

Similarly, for the channel $W_s$ we have the following: For any $\epsilon>0$ and $0< \beta <\frac{1}{2}$, there exist a large $N=2^n$ and a partition $B_0,B_1$ of $[1,N]$ such that for $\tau=0,1$ and $i\in B_{\tau}$, $\left|\bar{I}(W_{s,N}^{(i)}) -\tau \right|<\epsilon$ and such that for $i\in B_1$, $Z(W_{s,N}^{(i)})< 2^{-N^{\beta}}$. Moreover, as $\epsilon\rightarrow 0$ (and $N\rightarrow \infty$), $\frac{|B_\tau|}{N}\rightarrow q_\tau$ for some $q_0,q_1$ adding up to one with $q_1=\bar{I}(W_s)$.

\begin{lemma}\label{lemma:Zc_Zs_RD2}
For $i=1,\cdots,N$, $Z(W_{c,N}^{(i)})\ge Z(W_{s,N}^{(i)})$.
\end{lemma}
\begin{IEEEproof}
Follows from Lemma \ref{lemma:Wc_deg_Ws_RD2}, Lemma \ref{lemma:Degraded_ZZ} and Lemma \ref{lemma:Degraded_WW}.
\end{IEEEproof}

To introduce the encoding and decoding rules, we need to make the following definitions:
\begin{align*}
&A_0=\left\{i\in[1,N]\left|Z(W_{c,N}^{(i)})>2^{-N^{\beta}}\right.\right\}\\
&B_0=\left\{i\in[1,N]\left|Z(W_{s,N}^{(i)})>1-2^{-N^{\beta}}\right.\right\}
\end{align*}
and $A_1=[1,N]\backslash A_0$ and $B_1=[1,N]\backslash B_0$. For $t=0,1$ and $\tau=0,1$, define $A_{t,\tau}=A_t\cap B_{\tau}$. Note that for large $N$, $2^{-N^{\beta}} < 1-2^{-N^{\beta}}$ and therefore, Lemma \ref{lemma:Zc_Zs_RD2} implies $A_{1,0}=\emptyset$. Note that the above polarization results imply that as $N$ increases, $\frac{|A_1|}{N}\rightarrow \bar{I}(W_c)$ and $\frac{|B_{\tau}|}{N}\rightarrow \bar{I}(W_s)$.\\

\subsubsection{Encoding and Decoding}
Let $z_1^N\in\G^N$ be an outcome of the random variable $Z_1^N$ known to both the encoder and the decoder. Given a source sequence $x_1^N\in\mathcal{X}^N$, the encoding rule is as follows: For $i\in[1,N]$, if $i\in B_0$, then $v_i$ is uniformly distributed over $\G$ and is known to both the encoder and the decoder (and is independent from other random variables). If $i\in B_1$, $v_i=g$ for some $g\in\G$ with probability
\begin{align*}
P(v_i=g)=p_{V_i|X_1^NZ_1^NV_1^{i-1}}(g|x_1^N,z_1^N,v_1^{i-1})
\end{align*}
Note that $[1,N]$ can be partitioned into $A_{0,0},A_{0,1}$ and $A_{1,1}$ (since $A_{1,0}$ is empty) and $B_0=A_{0,0}$, $B_1=A_{0,1}\cup A_{1,1}$. Therefore, $v_-1^N$ can be decompose as $v_1^N=v_{A_{0,0}}+v_{A_{0,1}}+v_{A_{1,1}}$ in which $v_{A_{0,0}}$ is known to the decoder. The encoder sends $v_{A_{0,1}}$ to the decoder and the decoder uses the channel code to recover $v_{A_{1,1}}$. The decoding rule is as follows: Given $z_1^N$, $v_{A_{0,0}}$ and $v_{A_{0,1}}$, let $\hat{v}_{A_{0,0}}=v_{A_{0,0}}$ and $\hat{v}_{A_{0,1}}=v_{A_{0,1}}$. For $i\in A_{1,1}$, let
\begin{align*}
\hat{v}_i=\argmax_{g\in G} W_{c,N}^{(i)}(z_1^N,\hat{v}_1^{i-1}|g)
\end{align*}
Finally, the decoder outputs $z_1^N-\hat{v}_1^NG$.

\subsubsection{Error Analysis}
The analysis is a combination of the-point-to point channel coding and source coding results for polar codes. The average distortion between the encoder input and the decoder output is upper bounded by
\begin{align*}
D_{avg}&\!\le\!\!\!\!\!\sum_{z_1^N\in\G^N} \!\! \frac{1}{q^N} \!\!\!\!\!
\sum_{x_1^N\in\mathcal{X}^N} \!\!\!\!\! p_X^N(x_1^N)\!\!\!\!\!\!
\sum_{v_1^N\in\G^N} \!\! \frac{1}{q^{|B_0|}} \! \left(\prod_{i\in B_1}\!\!\! p(v_i|x_1^N\!\!,z_1^N\!\!,\!v_1^{i-1}\!)\!\!\right)\\
&\qquad\qquad\qquad\quad \Big(d_{max}\cdot \mathds{1}_{\{\hat{v}\ne v\}}+d(x_1^N,z_1^N-v_1^NG)\Big)
\end{align*}
where we have replaced $p_{V_i|X_1^NZ_1^NV_1^{i-1}}(v_i|x_1^N,z_1^N,v_1^{i-1})$ with $p(v_i|x_1^N,z_1^N,v_1^{i-1})$ for simplicity of notation and $d_{max}$ is the maximum value of the $d(\cdot,\cdot)$ function. Let
\begin{align*}
&q_{V_i|X_1^NZ_1^NV_1^{i-1}}(v_i|x_1^N\!z_1^N\!v_1^{i-1})\\
&\qquad\qquad\quad =\left\{\begin{array}{ll}
\!\!\!\frac{1}{2}& \mbox{ If } i\in B_0\\
\!\!\!p_{V_i|X_1^NZ_1^NV_1^{i-1}}(v_i|x_1^N\!z_1^N\!v_1^{i-1})& \mbox{ If }i\in B_1
\end{array}\right.
\end{align*}
and
\begin{align*}
q_{X_1^NZ_1^N}(x_1^N,z_1^N)=p_{X_1^NZ_1^N}(x_1^N,z_1^N)
\end{align*}
We have
\begin{align}\label{eqn:Davg}
\nonumber D_{avg}&\le\sum_{\substack{v_1^N,z_1^N\in\G^N\\x_1^N\in\mathcal{X}^N}} q_{V_1^NX_1^NZ_1^N}(v_1^N,x_1^N,z_1^N)\\
&\nonumber \qquad\qquad\qquad \Big(d_{max}\cdot \mathds{1}_{\{\hat{v}\ne v\}}+d(x_1^N,z_1^N-v_1^NG)\Big)\\
&\nonumber\le \!\!\!\!\!\!\!\!\!\!\sum_{\substack{v_1^N,z_1^N\in\G^N\\x_1^N\in\mathcal{X}^N}}\!\!\!\!\!\!\!\!\! \Big(p(v_1^N,x_1^N,z_1^N) \!+\! \left|q(v_1^N\!\!,x_1^N\!\!,z_1^N)\!-\!p(v_1^N\!\!,x_1^N\!\!,z_1^N)\right|\Big)\\
&\qquad\qquad \Big(d_{max}\cdot \mathds{1}_{\{\hat{v}\ne v\}}+d(x_1^N,z_1^N-v_1^NG)\Big)
\end{align}
where in the last inequality, we dropped the subscripts of the probability distributions for simplicity of notation.
Therefore,
\begin{align}\label{eqn:D1D2D3}
D_{avg}&\le D_1+D_2+D_3
\end{align}
where
\begin{align}
&\label{eqn:D1}D_{1} = \sum_{\substack{v_1^N,z_1^N\in\G^N\\x_1^N\in\mathcal{X}^N}} p(v_1^N,x_1^N,z_1^N) d_{max}\cdot \mathds{1}_{\{\hat{v}\ne v\}}\\
&\label{eqn:D2}D_{2} = \sum_{\substack{v_1^N,z_1^N\in\G^N\\x_1^N\in\mathcal{X}^N}} p(v_1^N,x_1^N,z_1^N) d(x_1^N,z_1^N-v_1^NG)\\
&\nonumber D_{3} = \sum_{\substack{v_1^N,z_1^N\in\G^N\\x_1^N\in\mathcal{X}^N}}  \left|q(v_1^N,x_1^N,z_1^N)-p(v_1^N,x_1^N,z_1^N)\right|\\
&\label{eqn:D3}\qquad\qquad\qquad \Big(d_{max}\cdot \mathds{1}_{\{\hat{v}\ne v\}}+d(x_1^N,z_1^N-v_1^NG)\Big)
\end{align}

Here, we only give a sketch for the rest of the proof. The proof for the general case is completely presented in Section \ref{section:General_proof_RD}. The proof proceeds as follows: It is straightforward to show that $D_1\rightarrow D$ as $N$ increases. It can also be shown that $D_2\rightarrow 0$ as $N$ increases since the inner code is a good channel code. Finally, it can be shown that $D_3\rightarrow 0$ as $N$ increases since the total variation distance between the $P$ and the $Q$ measures is small (in turn since the outer code is a good source code).
\subsection{Source Coding: Proof for the General Case}\label{section:General_proof_RD}
\subsubsection{Review of Polar Codes for Arbitrary Sources and Channels}
The result of channel polarization for arbitrary discrete memoryless channels applied to $W_c$ implies \cite{} that for any $\epsilon>0$ and $0< \beta <\frac{1}{2}$, there exist a large $N=2^n$ and a partition $\{A_H|H\le \G\}$ of $[1,N]$ such that for $H\le \G$ and $i\in A_H$, $\left|\bar{I}(W_{c,N}^{(i)}) -\log \frac{|\G|}{|H|} \right|<\epsilon$ and $Z^H(W_{c,N}^{(i)})< 2^{-N^{\beta}}$. Moreover, as $\epsilon\rightarrow 0$ (and $N\rightarrow \infty$), $\frac{|A_H|}{N}\rightarrow p_H$ for some probabilities $p_H,H\le \G$ adding up to one with $\sum_{H\le \G} p_H \log \frac{|\G|}{|H|}=\bar{I}(W_c)$.\\

Similarly, for the channel $W_s$ we have the following: For any $\epsilon>0$ and $0< \beta <\frac{1}{2}$, there exist a large $N=2^n$ and a partition $\{B_H|H\le \G\}$ of $[1,N]$ such that for $H\le \G$ and $i\in B_H$, $\left|\bar{I}(W_{s,N}^{(i)}) -\log \frac{|\G|}{|H|} \right|<\epsilon$ and $Z^H(W_{s,N}^{(i)})< 2^{-N^{\beta}}$. Moreover, as $\epsilon\rightarrow 0$ (and $N\rightarrow \infty$), $\frac{|B_H|}{N}\rightarrow q_H$ for some probabilities $q_H,H\le \G$ adding up to one with $\sum_{H\le \G} q_H \log \frac{|\G|}{|H|}=\bar{I}(W_s)$.

\begin{lemma}
If the channel $(\G,\mathcal{Y}_1,W_1)$ is degraded with respect to the channel $(\G,\mathcal{Y}_2,W_2)$ in the sense of Definition \ref{def:degraded}, then for any $d\in G$,
\begin{align*}
Z_d(W_1)\ge Z_d(W_2)
\end{align*}
\end{lemma}
\begin{IEEEproof}
Let $(\mathcal{Y}_2,\mathcal{Y}_1,W)$ be a channel so that the condition of Definition \ref{def:degraded} is satisfied. We have
\begin{align*}
Z_d(W_1)&=\frac{1}{q} \sum_{x\in\G} \sum_{y_1\in\mathcal{Y}_1} \sqrt{W_1(y_1|x)W_1(y_1|x+d)}\\
&=\frac{1}{q} \sum_{x\in\G} \sum_{y_1\in\mathcal{Y}_1} \\
&\sqrt{\!\sum_{y_2\in\mathcal{Y}_2} \!\!\!W_2(y_2|x) W\!(y_1|y_2)\!\!\!\!\sum_{y_2'\in\mathcal{Y}_2} \!\!\!W_2(y_2'|x+d) W\!(y_1|y_2')}\\
&\!\ge \!\frac{1}{q}\!\! \sum_{x\in\G} \sum_{y_1\in\mathcal{Y}_1} \sum_{y_2\in\mathcal{Y}_2} \!\!\!\!\!\sqrt{W_2(y_2|x) W\!(y_1|y_2)^2 W_2(y_2|x\!+\!d) }\\
&= \frac{1}{q} \sum_{x\in\G} \sum_{y_2\in\mathcal{Y}_2} \sqrt{W_2(y_2|x) W_2(y_2|x+d) }\\
&=Z_d(W_2)
\end{align*}

\end{IEEEproof}

\begin{lemma}\label{lemma:Zdc_Zds_RDG}
For $i=1,\cdots,N$ and for $d\in \G$ and $H\le\G$, $Z_d(W_{c,N}^{(i)})\ge Z_d(W_{s,N}^{(i)})$ and $Z^H(W_{c,N}^{(i)})\ge Z^H(W_{s,N}^{(i)})$.
\end{lemma}
\begin{IEEEproof}
Follows from Lemma \ref{lemma:Wc_deg_Ws_RD2}, Lemma \ref{lemma:Degraded_WW} and Lemma \ref{lemma:Zdc_Zds_RDG}.
\end{IEEEproof}
We define some quantities before we introduce the encoding and decoding rules. For $H\le \G$, define
\begin{align*}
&A_H=\Big\{i\in[1,N]\Big|Z^H(W_{c,N}^{(i)})<2^{-N^{\beta}},\\
&\qquad\qquad\qquad\qquad \nexists K\le H \mbox{such that } Z^K(W_{c,N}^{(i)})<2^{-N^{\beta}}\Big\}\\
&B_{H}=\Big\{i\in[1,N]\Big|Z^{H}(W_{s,N}^{(i)})<1-2^{-N^{\beta}},\\
&\qquad\qquad\qquad \nexists K\le H\mbox{ such that }Z^{K}(W_{s,N}^{(i)})<1-2^{-N^{\beta}}\Big\}
%&A_H=\left\{i\in[1,N]\left|Z^H(W_{c,N}^{(i)})<2^{-N^{\beta}}, \forall K\le H: Z^K(W_{c,N}^{(i)})>2^{-N^{\beta}}\right.\right\}\\
%&B_{H}=\left\{i\in[1,N]\left|Z^{H}(W_{s,N}^{(i)})<1-2^{-N^{\beta}}, \forall K\le H: Z^{K}(W_{c,N}^{(i)})>1-2^{-N^{\beta}}\right.\right\}
\end{align*}
For $H\le \G$ and $K\le \G$, define $A_{H,K}=A_H\cap B_{K}$. Note that for large $N$, $2^{-N^{\beta}} < 1-2^{-N^{\beta}}$ and therefore, if for some $i\in[1,N]$, $i\in A_H$, Lemma \ref{lemma:Zdc_Zds_RDG} implies $Z^H(W_{s,N}^{(i)})<1-2^{-N^{\beta}}$ and hence $i\in \cup_{K\le H}B_K$. Therefore, for $K \nleq H$, $A_{H,K}=\emptyset$. Therefore $\{A_{H,K}|K\le H\le \G\}$ is a partition of $[1,N]$. Note that the channel polarization results imply that as $N$ increases, $\frac{|A_H|}{N}\rightarrow p_H$ and $\frac{|B_{H}|}{N}\rightarrow q_H$.\\

\subsubsection{Encoding and Decoding}
Let $z_1^N\in\G^N$ be an outcome of the random variable $Z_1^N$ known to both the encoder and the decoder. Given $K\le H\le \G$, let $T_H$ be a transversal of $H$ in $\G$  and let $T_{K\le H}$ be a transversal of $K$ in $H$. Any element $g$ of $\G$ can be represented by $g=[g]_K+[g]_{T_{K\le H}}+[g]_{T_H}$ for unique $[g]_K\in K$, $[g]_{T_{K\le H}}\in T_{K\le H}$ and $[g]_{T_H}\in T_H$. Also note that $T_{K\le H}+T_H$ is a transversal $T_K$ of $K$ in $\G$ so that $g$ can be uniquely represented by $g=[g]_K+[g]_{T_K}$ for some $[g]_{T_K}\in T_K$ and $[g]_{T_K}$ can be uniquely represented by $[g]_{T_K}= [g]_{T_{K\le H}}+[g]_{T_H}$.

Given a source sequence $x_1^N\in\mathcal{X}^N$, the encoding rule is as follows: For $i\in[1,N]$, if $i\in A_{H,K}$ for some $K\le H\le \G$, $[v_i]_K$ is uniformly distributed over $K$ and is known to both the encoder and the decoder (and is independent from other random variables). The component $[v_i]_{T_K}$ is chosen randomly so that for $g\in [v_i]_K+T_K$,
\begin{align*}
P(v_i=g)=\frac{p_{V_i|X_1^NZ_1^NV_1^{i-1}}(g|x_1^N,z_1^N,v_1^{i-1})}{p_{V_i|X_1^NZ_1^NV_1^{i-1}}([v_i]_K+T_K|x_1^N,z_1^N,v_1^{i-1})}
\end{align*}
Note that $v_1^N$ can be decomposed as $v_1^N=[v_1^N]_K+[v_1^N]_{T_{K\le H}}+[v_1^N]_{T_H}$ (with a slight abuse of notation since $K$ and $H$ depend on the index $i$) in which $[v_1^N]_K$ is known to the decoder. The encoder sends $[v_1^N]_{T_{K\le H}}$ to the decoder and the decoder uses the channel code to recover $[v_1^N]_{T_H}$. The decoding rule is as follows: Given $z_1^N$, $[v_1^N]_K$ and $[v_1^N]_{T_{K\le H}}$, and for $i\in A_{H,K}$, let
\begin{align*}
\hat{v}_i=\argmax_{g\in [v_i]_K+[v_i]_{T_{K\le H}}+T_H} W_{c,N}^{(i)}(z_1^N,\hat{v}_1^{i-1}|g)
\end{align*}
Finally, the decoder outputs $z_1^N-\hat{v}_1^NG$. Note that the rate of this code is equal to
\begin{align*}
R &=\sum_{K\le H\le \G} \frac{|A_{H,K}|}{N} \log \frac{|H|}{|K|}\\
&= \sum_{K\le H\le \G} \frac{|A_{H,K}|}{N} \log \frac{|\G|}{|K|} - \sum_{K\le H\le \G} \frac{|A_{H,K}|}{N} \log \frac{|\G|}{|H|}\\
&= \sum_{K\le \G} \frac{|B_{K}|}{N} \log \frac{|\G|}{|K|} - \sum_{H\le \G} \frac{|A_{H}|}{N} \log \frac{|\G|}{|H|}\\
&\rightarrow \bar{I}(W_s)-\bar{I}(W_c)= I(X;U)
\end{align*}

\subsection{Error Analysis}
The average distortion between the encoder input and the decoder output is upper bounded by
\begin{align*}
D_{avg}&\le\sum_{z_1^N\in\G^N}\frac{1}{q^N}
\sum_{x_1^N\in\mathcal{X}^N}p_X^N(x_1^N)
\sum_{v_1^N\in\G^N} \frac{1}{q^{|B_0|}} \\
&\qquad\quad \left(\prod_{K\le \G} \prod_{i\in B_K} \!\! \frac{p(g|x_1^N,z_1^N,v_1^{i-1})}{p([v_i]_K+T_K|x_1^N,z_1^N,v_1^{i-1}) \cdot |K|}\right)\\
&\qquad\qquad\qquad\qquad \left(d_{max}\cdot \mathds{1}_{\{\hat{v}\ne v\}}+d(x_1^N,z_1^N-v_1^NG)\right)
\end{align*}
where $p_{V_i|X_1^NZ_1^NV_1^{i-1}}(\cdot|x_1^N,z_1^N,v_1^{i-1})$ is replaced with $p(\cdot|x_1^N,z_1^N,v_1^{i-1})$ for simplicity of notation. For $i\in B_K$, let
\begin{align*}
&q_{V_i|X_1^NZ_1^NV_1^{i-1}}(v_i|x_1^Nz_1^Nv_1^{i-1})\\
&\qquad = \frac{p_{V_i|X_1^NZ_1^NV_1^{i-1}}(g|x_1^N,z_1^N,v_1^{i-1})}{p_{V_i|X_1^NZ_1^NV_1^{i-1}}([v_i]_K+T_K|x_1^N,z_1^N,v_1^{i-1})\cdot |K|}
\end{align*}
and
\begin{align*}
q_{X_1^NZ_1^N}(x_1^N,z_1^N)=p_{X_1^NZ_1^N}(x_1^N,z_1^N)
\end{align*}
Note that Equations \eqref{eqn:Davg} through \eqref{eqn:D3} are valid for the general case (with the new $Q$ measure).

%We have
%\begin{align*}
%D_{avg}&\le\sum_{\substack{v_1^N,z_1^N\in\G^N\\x_1^N\in\mathcal{X}^N}} q_{V_1^NX_1^NZ_1^N}(v_1^N,x_1^N,z_1^N) \left(d_{max}\cdot \mathds{1}_{\{\hat{v}\ne v\}}+d(x_1^N,z_1^N-v_1^NG)\right)\\
%&\le \sum_{\substack{v_1^N,z_1^N\in\G^N\\x_1^N\in\mathcal{X}^N}} \left(p(v_1^N,x_1^N,z_1^N) +\left|q(v_1^N,x_1^N,z_1^N)-p(v_1^N,x_1^N,z_1^N)\right|\right) \left(d_{max}\cdot \mathds{1}_{\{\hat{v}\ne v\}}+d(x_1^N,z_1^N-v_1^NG)\right)
%\end{align*}
%where in the last inequality, we dropped the subscripts of the probability distributions for simplicity of notation.
%Therefore,
%\begin{align*}
%D_{avg}&\le D_1+D_2+D_3
%\end{align*}
%where
%\begin{align*}
%&D_{1} = \sum_{\substack{v_1^N,z_1^N\in\G^N\\x_1^N\in\mathcal{X}^N}} p(v_1^N,x_1^N,z_1^N) d_{max}\cdot \mathds{1}_{\{\hat{v}\ne v\}}\\
%&D_{2} = \sum_{\substack{v_1^N,z_1^N\in\G^N\\x_1^N\in\mathcal{X}^N}} p(v_1^N,x_1^N,z_1^N) d(x_1^N,z_1^N-v_1^NG)\\
%&D_{3} = \sum_{\substack{v_1^N,z_1^N\in\G^N\\x_1^N\in\mathcal{X}^N}}  \left|q(v_1^N,x_1^N,z_1^N)-p(v_1^N,x_1^N,z_1^N)\right| \left(d_{max}\cdot \mathds{1}_{\{\hat{v}\ne v\}}+d(x_1^N,z_1^N-v_1^NG)\right)
%\end{align*}

It follows from the analysis of \cite[Section III.F and Section IV]{sahebi_polar_source} that $D_2=D$. The following lemma is also proved in \cite[Section III.F and Section IV]{sahebi_polar_source}.

\begin{lemma}
With the above definitions,
\begin{align*}
\|P-Q\|_{t.v.} &= \sum_{\substack{v_1^N,z_1^N\in\G^N\\x_1^N\in\mathcal{X}^N}}  \left|q(v_1^N,x_1^N,z_1^N)-p(v_1^N,x_1^N,z_1^N)\right|\\
&\le K2^{-N^{\beta}}
\end{align*}
for some constant $K$ depending only on $q$.
\end{lemma}

It remains to show that $D_1$ vanishes as $N$ approaches infinity. We have
\begin{align*}%D_1 &=d_{max}
&\sum_{\substack{v_1^N\!\!,z_1^N\in\G^N\\x_1^N\in\mathcal{X}^N}} \!\!\!\!\!\!\! p_{XU}^N(x_1^N\!\!,z_1^N\!\!-v_1^NG) \!\!\!\!\!\! \sum_{K\le H\le \G} \sum_{i\in A_{H,K}} \mathds{1}_{\Big\{W_{c,N}^{(i)}(z_1^N,v_1^{i-1}|v_i)}\\
&\quad {\color{white}1}_{\le W_{c,N}^{(i)}(z_1^N,v_1^{i-1}|\tilde{v}_i) \mbox{ for some }\tilde{v}_i\in [v_i]_K+[v_i]_{T_{K\le H}}+T_H,\tilde{v}_i\ne v_i\Big\}}\\
&\le \sum_{\substack{v_1^N,z_1^N\in\G^N\\x_1^N\in\mathcal{X}^N}} \!\!\!\!\!\!p_{XU}^N(x_1^N,z_1^N-v_1^NG) \!\!\!\!\sum_{K\le H\le \G} \sum_{i\in A_{H,K}} \sum_{\substack{\tilde{v}_i\in [v_i]_H+T_H\\\tilde{v}_i\ne v_i}}\\
&\qquad\qquad\qquad\qquad\qquad\qquad\qquad \sqrt{\frac{W_{c,N}^{(i)}(z_1^N,v_1^{i-1}|\tilde{v}_i)}{W_{c,N}^{(i)}(z_1^N,v_1^{i-1}|v_i)}}\\
&=\!\!\!\!\!\! \sum_{K\le H\le \G} \sum_{i\in A_{H,K}} \!\!\!\!\sum_{\substack{v_i\in \G\\\tilde{v}_i\in [v_i]_H+T_H\\\tilde{v}_i\ne v_i}} \!\!\sum_{v_1^{i-1},z_1^N} \!\!\!\frac{1}{q} \!\!\left(\sum_{v_{i+1^N}} \!\!\!\!\frac{1}{q^{N-1}} p_{U}^N(z_1^N\!\!\!-\!v_1^NG)\!\!\right)\\
&\qquad\qquad\qquad\qquad\qquad\qquad\qquad \sqrt{\frac{W_{c,N}^{(i)}(z_1^N,v_1^{i-1}|\tilde{v}_i)}{W_{c,N}^{(i)}(z_1^N,v_1^{i-1}|v_i)}}\\
&= \sum_{K\le H\le \G} \sum_{i\in A_{H,K}} \sum_{\substack{v_i\in \G\\\tilde{v}_i\in [v_i]_H+T_H\\\tilde{v}_i\ne v_i}} Z_{\{v_i,\tilde{v}_i\}}(W_{c,N}^{(i)})
\end{align*}
Note that $\tilde{v}_i\in [v_i]_H+T_H$ and $\tilde{v}_i\ne v_i$ imply $d=\tilde{v}_i-v_i\notin H$. We have
\begin{align*}
Z_{\{v_i,\tilde{v}_i\}}(W_{c,N}^{(i)})\le qZ_d(W_{c,N}^{(i)})\le q Z^H(W_{c,N}^{(i)})
\end{align*}
Therefore,
\begin{align*}
D_1 &\le \sum_{K\le H\le \G} \sum_{i\in A_{H,K}} \sum_{\substack{v_i\in \G\\\tilde{v}_i\in [v_i]_H+T_H\\\tilde{v}_i\ne v_i}} q Z^H(W_{c,N}^{(i)})\\
&\le 4^q q N 2^{-N^{\beta}}
\end{align*}
Therefore, $D_1\rightarrow 0$ as $N$ increases.

\section{Polar Codes Achieve the Shannon Capacity of Arbitrary DMCs}
In this section, we prove the following theorem:
\begin{theorem}
For an arbitrary discrete memoryless channel $(\mathcal{X},\mathcal{Y},W)$, nested polar codes achieve the Shannon capacity.
\end{theorem}

For the channel let $\mathcal{X}=\G$ for some Abelian group $\G$ and let $|\G|=q$. Similarly to the source coding problem, we show that there exists nested polar code $\mathds{C}_i\subseteq \mathcal{C}_o$ such that $\mathds{C}_o$ is a good channel code and each shift of $\mathds{C}_i$ is a good source code. This will be made clear later in the following.\\

\begin{figure}[!h]\label{fig:Ws_C}
\centering
\includegraphics[scale=1]{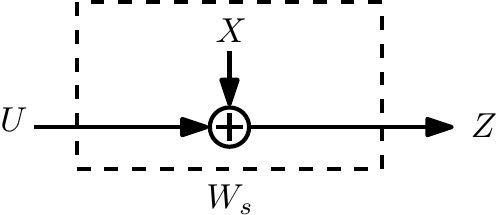}
\caption{\small Test channel for the inner code (the source coding component)}
\end{figure}

\begin{figure}[!h]\label{fig:Wc_C}
\centering
\includegraphics[scale=1]{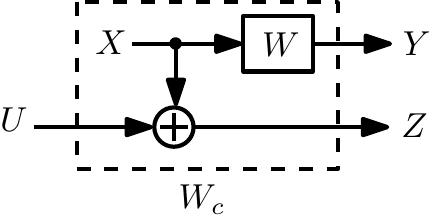}
\caption{\small Test channel for the outer code (the channel coding component)}
\end{figure}

Let $X$ be a random variable with the capacity achieving distribution and let $U$ be uniformly distributed over $\G$. Define the artificial channels $(\G,\G,W_s)$ and $(\G,\mathcal{Y}\times\G,W_c)$ such that for $u,z\in \G$ and $y\in \mathcal{Y}$,
\begin{align*}
&W_s(z|u)=p_X(z-u)\\
&W_c(y,z|u)=p_{XY}(z-u,y)
\end{align*}
These channels have been depicted in Figures \ref{fig:Ws_C} and \ref{fig:Wc_C}.

Note that for $u,x,z\in G$ and $y\in\mathcal{Y}$, $p_{UXYZ}(u,x,y,z)=p_U(u)p_X(x)W(y|x)\mathds{1}_{\{z=u+x\}}$. Similarly to the source coding case, one can show that the symmetric capacities of the channels are equal to
\begin{align*}
&\bar{I}(W_s)=\log q-H(X)\\
&\bar{I}(W_c)=\log q-H(X|Y)
\end{align*}

We employ a nested polar code in which the inner code is a good source code for the test channel $W_s$ and the outer code is a good channel code for $W_c$. The rate of this code is equal to
\begin{align*}
R&=\bar{I}(W_c)-\bar{I}(W_x)\\
&=\log q-H(X|Y)-\left(\log q-H(X)\right)=I(X;Y)
\end{align*}
Note that the channels $W_c$ and $W_s$ are chosen so that the difference of their \emph{symmetric} capacities is equal to the \emph{Shannon} capacity of the original channel. We postpone the proof to Section \ref{section:proof_C2} where the result is proved for the binary case and Section \ref{section:proof_CG} in which the general proof (for arbitrary Abelian groups) is presented. The rest of this section is devoted to some general definitions and lemmas which are used in the proofs.

Let $n$ be a positive integer and let $N=2^n$. Similar to the source coding case, For both channels $W_s$ and $W_c$ and for $i=1\cdots N$, define the synthesized channels as
\begin{align*}
W_{c,N}^{(i)}(y_1^N,z_1^N,v_1^{i-1}|v_i)&=\sum_{v_{i+1}^N\in G^{N-i}} \frac{1}{2^{N-1}} W_c^N(y_1^N,z_1^N|v_1^NG)\\
&=\sum_{v_{i+1}^N\in G^{N-i}} \frac{1}{2^{N-1}} p_{XY}^N(z_1^N-v_1^NG,y_1^N)
\end{align*}
and
\begin{align*}
W_{s,N}^{(i)}(z_1^N,v_1^{i-1}|v_i)&=\sum_{v_{i+1}^N\in G^{N-i}} \frac{1}{2^{N-1}} W_s^N(z_1^N|v_1^NG)\\
&=\sum_{v_{i+1}^N\in G^{N-i}} \frac{1}{2^{N-1}} p_X^N(z_1^N-v_1^NG)
\end{align*}

%Recall that
%\begin{align*}
%Z(W_{c,N}^{(i)})=\sum_{y_1^N,z_1^N} \sum_{v_1^{i-1}} \sqrt{W_{c,N}^{(i)}(y_1^N,z_1^N,v_1^{i-1}|0) W_{c,N}^{(i)}(y_1^N,z_1^N,v_1^{i-1}|1)}
%\end{align*}
%and
%\begin{align*}
%Z(W_{s,N}^{(i)})=\sum_{z_1^N} \sum_{v_1^{i-1}} \sqrt{W_{s,N}^{(i)}(z_1^N,v_1^{i-1}|0) W_{s,N}^{(i)}(z_1^N,v_1^{i-1}|1)}
%\end{align*}

Let the random vector $U_1^N$ be distributed according to $p_U^N$ (uniform) and let $V_1^N=U_1^N G^{-1}$ where $G$ is the polar coding matrix of dimension $N\times N$. Note that since $G$ is a one-to-one mapping, $V_1^N$ is also uniformly distributed. Let $Y_1^N$ and $Z_1^N$ be the outputs of the channel $W_c$ when the input is $U_1^N$. Note that for $v_1^N,u_1^n,x_1^N,z_1^N\in \G^N$ and $y_1^N\in \mathcal{Y}^N$,
\begin{align*}
&p_{V_1^N U_1^N X_1^N Y_1^N Z_1^N}(v_1^N,u_1^n,x_1^N,y_1^N,z_1^N)\\
&=\mathds{1}_{\{v_1^N=u_1^N G^{-1}\}} p_U^N(u_1^N) p_X^N(x_1^N) W^N(y_1^N|x_1^N) \mathds{1}_{\{z_1^N=u_1^N+x_1^N\}}\\
&=\frac{1}{2^N} p_X^N(z_1^N-v_1^N G) W^N(y_1^N|z_1^N-v_1^N G) \mathds{1}_{\{u_1^N=v_1^N G,x_1^N =z_1^N-v_1^NG\}}
\end{align*}
and
\begin{align*}
p_{V_1^N Y_1^N Z_1^N}(v_1^N,y_1^N,z_1^N) &=\frac{1}{2^N} p_X^N(z_1^N-v_1^N G) W^N(y_1^N|z_1^N-v_1^N G)\\
p_{V_1^N Z_1^N}(v_1^N,z_1^N) &=\frac{1}{2^N} p_X^N(z_1^N-v_1^N G)
\end{align*}

\section{Channel Coding: Sketch of the Proof for the Binary Case}\label{section:proof_C2}
The following theorems state the standard channel coding and source coding polarization phenomenons for the binary case.
\begin{theorem}\label{theorem:polar_source2C}
For any $\epsilon>0$ and $0< \beta <\frac{1}{2}$, there exist a large $N=2^n$ and a partition $A_0,A_1$ of $[1,N]$ such that for $t=0,1$ and $i\in A_t$, $\left|\bar{I}(W_{s,N}^{(i)}) -t \right|<\epsilon$ and $Z(W_{s,N}^{(i)})< 2^{-N^{\beta}}$. Moreover, as $\epsilon\rightarrow 0$ (and $N\rightarrow \infty$), $\frac{|A_t|}{N}\rightarrow p_t$ for some $p_0,p_1$ adding up to one with $p_1=\bar{I}(W_s)$.
\end{theorem}

\begin{theorem}\label{theorem:polar_channel2C}
For any $\epsilon>0$ and $0< \beta <\frac{1}{2}$, there exist a large $N=2^n$ and a partition $B_0,B_1$ of $[1,N]$ such that for $\tau=0,1$ and $i\in B_{\tau}$, $\left|\bar{I}(W_{c,N}^{(i)}) -\tau \right|<\epsilon$ and $Z(W_{c,N}^{(i)})< 2^{-N^{\beta}}$. Moreover, as $\epsilon\rightarrow 0$ (and $N\rightarrow \infty$), $\frac{|B_\tau|}{N}\rightarrow q_\tau$ for some $q_0,q_1$ adding up to one with $q_1=\bar{I}(W_c)$.
\end{theorem}

\begin{lemma}\label{lemma:Zc_Zs_C2}
For $i=1,\cdots,N$, $Z(W_{s,N}^{(i)})\ge Z(W_{c,N}^{(i)})$.
\end{lemma}
\begin{IEEEproof}
Follows from since $W_s$ is degraded with respect to $W_c$ and using Lemma[???] and Lemma[???].
\end{IEEEproof}

Define
\begin{align*}
&A_0=\left\{i\in[1,N]\left|Z(W_{s,N}^{(i)})>1-2^{-N^{\beta}}\right.\right\}\\
&B_0=\left\{i\in[1,N]\left|Z(W_{c,N}^{(i)})>2^{-N^{\beta}}\right.\right\}
\end{align*}
and $A_1=[1,N]\backslash A_0$ and $B_1=[1,N]\backslash B_0$. For $t=0,1$ and $\tau=0,1$, define $A_{t,\tau}=A_t\cap B_{\tau}$. Note that Lemma \ref{lemma:Zc_Zs_RD2} implies
\begin{align*}
A_{1,0}=\left\{i\in[1,N]\left|<2^{-N^{\beta}}<Z(W_{c,N}^{(i)})<Z(W_{s,N}^{(i)})<1-2^{-N^{\beta}}\right.\right\}
\end{align*}
Since $Z(W_{c,N}^{(i)})$ and $Z(W_{s,N}^{(i)})$ both polarize to $0,1$, as $N$ increases $\frac{A_{1,0}}{N}\rightarrow 0$.
Note that Theorems \ref{theorem:polar_channel2} and $\ref{theorem:polar_source2}$ imply that as $N$ increases, $\frac{|A_1|}{N}\rightarrow \bar{I}(W_c)$ and $\frac{|B_{1}|}{N}\rightarrow \bar{I}(W_s)$.

\subsection{Encoding and Decoding}
Let $z_1^N$ be a realization of the random vector $Z_1^N$ available to both the encoder and the decoder. Given a partition $A_{0,0},A_{0,1},A_{1,0},A_{1,1}$ of $[1,N]$, a vector $v_1^N\in \G^N$ can be decomposed as $v_1^N=v_{A_{0,0}}+v_{A_{0,1}}+v_{A_{1,0}}+v_{A_{1,1}}$ and similarly, the set $\G^N$ can be partitioned into the union $\G^{A_{0,0}}\cup\G^{A_{0,1}}\cup\G^{A_{1,0}}\cup\G^{A_{1,1}}$. Let $v_{A_{0,0}}\in \G^{A_{0,0}}$ be a uniformly distributed random variable available to both the encoder and the decoder which is independent from all other random variables and let $v_{A_{0,1}}$ be the message vector. The encoding is as follows: For $i \in A_{1,1}\cup A_{1,0}$,
\begin{align*}
v_i=\left\{\begin{array}{ll}
0& \mbox{with prob. } p_{V_i|Z_1^N V_1^{i-1}}(0|z_1^N,v_1^{i-1})\\
1& \mbox{with prob. } p_{V_i|Z_1^N V_1^{i-1}}(1|z_1^N,v_1^{i-1})
\end{array}\right.
\end{align*}

The receiver has access to $y_1^N$, $z_1^N$ and $v_{A_{0,0}}$. Note that $\frac{|A_{1,0}|}{N}\rightarrow 0$ as $N$ increases. Assume for the moment that the receiver has access to $v_{A_{1,0}}$. Then it can use the following decoding rule: For $i\in A_{0,1}\cup A_{1,1}$,
\begin{align*}
\hat{v}_i=\argmax_{g\in \G} W_{c,N}^{(i)} (y_1^N,z_1^N,\hat{v}_1^{i-1}|g)
\end{align*}
It is shown in the next section that with this encoding and decoding rules, the probability of error goes to zero. It remains to send the component $v_{A_{1,0}}$ to the decoder which can be done using a regular polar code (which achieves the symmetric capacity of the channel). Note that since the fraction $\frac{|A_{1,0}|}{N}$ vanishes as $N$ increases, the rate loss due to the transmission of $v_{A_{1,0}}$ can be made arbitrarily small.

\subsection{Error Analysis}
%Given the message $M_k$, an encoding error occurs if for all $|M_k|$ trials, $v_{A_{1,0}}\ne \tilde{v}_{A_{1,0}}$. Note that given the value of $v_{A_{0,0}}$, $v_{A_{1,0}}$'s and $\tilde{v}_{A_{1,0}}$ are independent and $\tilde{v}_{A_{1,0}}$ is uniformly distributed. Define
%\begin{align*}
%\theta(M_k,v_{A_{0,0}},z_1^N)=\sum_{v_{A_{0,1}}\in M_k} \mathds{1}_{\{v_{A_{1,0}}(v_{A_{0,0}},v_{A_{0,1}})=\tilde{v}_{A_{1,0}}\}}
%\end{align*}
%
%Therefore, the probability of encoding error given the message $M_k$ and the value of $v_{A_{0,0}}$ is given by
%\begin{align*}
%P(err|M_k,v_{A_{0,0}})&=\left(1-\frac{1}{|\G|^{|A_{1,0}|}}\right)^{|M_k|}\\
%&\le \left(1-\frac{1}{|\G|^{|A_{1,0}|}}\right)^{|\G|^{|A_{1,0}|+N\epsilon}}\\
%&\rightarrow e^{-|\G|^{N\epsilon}}\rightarrow 0
%\end{align*}
%If there is no encoding error, the receiver has access to $z_1^N, v_{A_{0,0}}, v_{A_{1,0}}$ and $y_1^N$. A communication error occurs if $\hat{v}_{0,1}\ne v_{0,1}$. The error event is contained in the event:
The receiver has access to $z_1^N, v_{A_{0,0}}, v_{A_{1,0}}$ and $y_1^N$. A communication error occurs if $\hat{v}_{0,1}\ne v_{0,1}$. The error event is contained in the event:
\begin{align*}
\bigcup_{i\in A_{0,1}\cup A_{1,1}} \{W_{c,N}^{(i)} (y_1^N,z_1^N,\hat{v}_1^{i-1}|v_i)\le W_{c,N}^{(i)} (y_1^N,z_1^N,\hat{v}_1^{i-1}|v_i+1)\}
\end{align*}

Therefore, we have the following upper bound on the average probability of error:
\begin{align*}
\mathds{E}\{P_{err}\}&\le \sum_{z_1^N\in\G^N} \frac{1}{q^N} \sum_{v_1^N\in\G^N} \frac{1}{q^{|A_{0,0}|+|A_{0,1}|}} \\ &\left(\prod_{i\in A_{0,1}\cup A_{1,0}} p_{V_i|Z_1^NV_1^{i-1}}(v_i|z_1^N,v_1^{i-1})\right) \sum_{y_1^N}W^N(y_1^N|z_1^N-v_1^NG) \\
&\mathds{1}_{\{\exists i\in[1,N]:W_{c,N}^{(i)} (y_1^N,z_1^N,{v}_1^{i-1}|{v}_i)\le W_{c,N}^{(i)} (y_1^N,z_1^N,{v}_1^{i-1}|{v}_i+1)\}}
%&\le \sum_{z_1^N} \frac{1}{2^N} \sum_{v_m} \frac{1}{2^{k_m}} \sum_{v_c,v_d}\frac{1}{2^{k_d}} \left(\prod_{i\in A_s} p_{V_i|Z_1^NV_1^{i-1}}(v_i|z_1^N,v_1^{i-1})\right) \sum_{y_1^N}W^N(y_1^N|z_1^N-v_1^NG) \\
%&\sum_{i\in A_m}\mathds{1}_{\{W_{c,N}^{(i)} (y_1^N,z_1^N,\hat{v}_1^{i-1}|\hat{v}_i)\le W_{c,N}^{(i)} (y_1^N,z_1^N,\hat{v}_1^{i-1}|\hat{v}_i+1)\}}
\end{align*}
Define
\begin{align*}
p(v_1^N,z_1^N)&=p_{V_1^NZ_1^N}(v_1^N,z_1^N)\\
&=\frac{1}{q^N}\cdot \left(\prod_{i\in [1,N]} p_{V_i|Z_1^NV_1^{i-1}}(v_i|z_1^N,v_1^{i-1})\right)
\end{align*}
and
\begin{align*}
q(v_1^N,z_1^N)=\frac{1}{q^N}\cdot \frac{1}{q^{|A_{0,0}|+|A_{0,1}|}} \left(\prod_{i\in A_{0,1}\cup A_{1,0}} p_{V_i|Z_1^NV_1^{i-1}}(v_i|z_1^N,v_1^{i-1})\right)
\end{align*}

Note that
\begin{align*}
\mathds{E}\{P_{err}\}&\le \sum_{v_1^N} \sum_{z_1^N} q(v_1^N,z_1^N) \sum_{y_1^N}W^N(y_1^N|z_1^N-v_1^NG)\\
& \mathds{1}_{\{\exists i\in[1,N]:W_{c,N}^{(i)} (y_1^N,z_1^N,{v}_1^{i-1}|{v}_i)\le W_{c,N}^{(i)} (y_1^N,z_1^N,{v}_1^{i-1}|{v}_i+1)\}}\\
&\le P_1+P_2
\end{align*}
where
\begin{align*}
P_1&= \sum_{v_1^N} \sum_{z_1^N} p(v_1^N,z_1^N) \sum_{y_1^N}W^N(y_1^N|z_1^N-v_1^NG)\\
&\mathds{1}_{\{\exists i\in[1,N]:W_{c,N}^{(i)} (y_1^N,z_1^N,{v}_1^{i-1}|{v}_i)\le W_{c,N}^{(i)} (y_1^N,z_1^N,{v}_1^{i-1}|{v}_i+1)\}}
\end{align*}
and
\begin{align*}
P_2&= \sum_{v_1^N} \sum_{z_1^N} |q(v_1^N,z_1^N)-p(v_1^N,z_1^N)| \sum_{y_1^N}W^N(y_1^N|z_1^N-v_1^NG) \\
&\mathds{1}_{\{\exists i\in[1,N]:W_{c,N}^{(i)} (y_1^N,z_1^N,{v}_1^{i-1}|{v}_i)\le W_{c,N}^{(i)} (y_1^N,z_1^N,{v}_1^{i-1}|{v}_i+1)\}}\\
&\le \sum_{v_1^N} \sum_{z_1^N} |q(v_1^N,z_1^N)-p(v_1^N,z_1^N)| \sum_{y_1^N}W^N(y_1^N|z_1^N-v_1^NG)\\
&\le \sum_{v_1^N} \sum_{z_1^N} |q(v_1^N,z_1^N)-p(v_1^N,z_1^N)|
\end{align*}
We use the following two lemmas from [???].

\begin{lemma}
For $p(\cdot,\cdot)$ and $q(\cdot,\cdot)$ defined as above,
\begin{align*}
&\sum_{v_1^N}\sum_{z_1^N} \left|p(v_1^N,z_1^N)-q(v_1^N,z_1^N)\right|\le 2\sum_{i\in A_{0,0}\cup A_{0,1}}\\
&\mathds{E}\left\{\left|\frac{1}{2}-p_{V_i|V_1^{i-1}Z_1^N}(0|V_1^N,Z_1^N)\right|\right\}
\end{align*}
\end{lemma}

\begin{lemma}
For $i\in [1,N]$, if $Z(W_{s,N}^{(i)})\ge 1-\delta_N^2$ then
\begin{align*}
\mathds{E}\left\{\left|\frac{1}{2}-p_{V_i|V_1^{i-1}Z_1^N}(0|V_1^N,Z_1^N)\right|\right\}\le \sqrt{2}\delta_N
\end{align*}
\end{lemma}
Note that for $i\in A_{0,0}\cup A_{0,1}$, $Z(W_{s,N}^{(i)})\ge 1-\delta_N^2$. Therefore, Lemmas [] and [] imply
\begin{align*}
P_2\le 2\sqrt{2}N\delta_N
\end{align*}
We have
\begin{align*}
P_1&\le \sum_{v_1^N} \sum_{z_1^N} p(v_1^N,z_1^N) \sum_{y_1^N}W^N(y_1^N|z_1^N-v_1^NG) \sum_{i\in A_{0,1}}\\
&\mathds{1}_{\{W_{c,N}^{(i)} (y_1^N,z_1^N,{v}_1^{i-1}|{v}_i)\le W_{c,N}^{(i)} (y_1^N,z_1^N,{v}_1^{i-1}|{v}_i+1)\}}\\
&\le \sum_{v_1^N} \sum_{z_1^N} p(v_1^N,z_1^N) \sum_{y_1^N}W^N(y_1^N|z_1^N-v_1^NG) \sum_{i\in A_{0,1}}\\
&\sqrt{\frac{W_{c,N}^{(i)} (y_1^N,z_1^N,{v}_1^{i-1}|{v}_i+1)}{W_{c,N}^{(i)} (y_1^N,z_1^N,{v}_1^{i-1}|{v}_i)}}\\
&= \sum_{v_1^N} \sum_{z_1^N} \sum_{y_1^N} \frac{1}{2^N} p_X^N(z_1^N-v_1^N G) W^N(y_1^N|z_1^N-v_1^NG) \\
&\sum_{i\in A_{0,1}}\sqrt{\frac{W_{c,N}^{(i)} (y_1^N,z_1^N,{v}_1^{i-1}|{v}_i+1)}{W_{c,N}^{(i)} (y_1^N,z_1^N,{v}_1^{i-1}|{v}_i)}}\\
\end{align*}
where the last equality follows since $p(v_1^N,z_1^N)=p_{V_1^NZ_1^N}(v_1^N,z_1^N)=\frac{1}{2^N} p_X^N(z_1^N-v_1^N G)$. Therefore,
\begin{align*}
P_1&\le \sum_{i\in A_{0,1}} \frac{1}{2} \sum_{v_i} \sum_{v_1^{i-1}} \sum_{y_1^N,z_1^N} \sqrt{\frac{W_{c,N}^{(i)} (y_1^N,z_1^N,{v}_1^{i-1}|{v}_i+1)}{W_{c,N}^{(i)} (y_1^N,z_1^N,{v}_1^{i-1}|{v}_i)}} \sum_{v_{i+1}^{N}} \frac{1}{2^{N-1}} p_X^N(z_1^N-v_1^N G) W^N(y_1^N|z_1^N-v_1^NG)\\
&=\sum_{i\in A_{0,1}} \frac{1}{2} \sum_{v_i} \sum_{v_1^{i-1}} \sum_{y_1^N,z_1^N} \sqrt{\frac{W_{c,N}^{(i)} (y_1^N,z_1^N,{v}_1^{i-1}|{v}_i+1)}{W_{c,N}^{(i)} (y_1^N,z_1^N,{v}_1^{i-1}|{v}_i)}} W_{c,N}^{(i)} (y_1^N,z_1^N,{v}_1^{i-1}|{v}_i)\\
&=\sum_{i\in A_{0,1}} \frac{1}{2} \sum_{v_i} Z(W_{c,N}^{(i)})\\
&\le N\delta_N^2
\end{align*}

\section{Channel Coding: Proof for the General Case}\label{section:proof_CG}
For $H\le \G$, define
\begin{align*}
&A_H=\Big\{i\in[1,N]\Big|Z^H(W_{s,N}^{(i)})<1-2^{-N^{\beta}},\\
&\qquad\qquad\qquad\qquad \nexists K\le H \mbox{such that } Z^K(W_{s,N}^{(i)})<1-2^{-N^{\beta}}\Big\}\\
&B_{H}=\Big\{i\in[1,N]\Big|Z^{H}(W_{c,N}^{(i)})<2^{-N^{\beta}},\\
&\qquad\qquad\qquad \nexists K\le H\mbox{ such that }Z^{K}(W_{c,N}^{(i)})<2^{-N^{\beta}}\Big\}
%&A_H=\left\{i\in[1,N]\left|Z^H(W_{c,N}^{(i)})<2^{-N^{\beta}}, \forall K\le H: Z^K(W_{c,N}^{(i)})>2^{-N^{\beta}}\right.\right\}\\
%&B_{H}=\left\{i\in[1,N]\left|Z^{H}(W_{s,N}^{(i)})<1-2^{-N^{\beta}}, \forall K\le H: Z^{K}(W_{c,N}^{(i)})>1-2^{-N^{\beta}}\right.\right\}
\end{align*}
For $H\le \G$ and $K\le \G$, define $A_{H,K}=A_H\cap B_{K}$. Note that for $K\le H\le \G$, $Z^H(W)\le Z^K(W)$. Also note that $Z^H(W_{c,N}^{(i)})\le Z^H(W_{s,N}^{(i)})$ Therefore, if $K\nleq H$ then
\begin{align*}
A_{H,K}\subseteq \Big\{i\in[1,N]\Big| 2^{-N^{\beta}}<Z^H(W_{c,N}^{(i)})\le Z^H(W_{s,N}^{(i)})<1-2^{-N^{\beta}}\Big\}
\end{align*}
Since $Z^H(W_{c,N}^{(i)})$ and $Z^H(W_{s,N}^{(i)})$ both polarize to $0,1$, as $N$ increases $\frac{A_{H,K}}{N}\rightarrow 0$ if $K\nleq H$. Note that the channel polarization results imply that as $N$ increases, $\frac{|A_H|}{N}\rightarrow p_H$ and $\frac{|B_{H}|}{N}\rightarrow q_H$.\\

\subsubsection{Encoding and Decoding}
Let $z_1^N\in\G^N$ be an outcome of the random variable $Z_1^N$ known to both the encoder and the decoder. Given $K\le H\le \G$, let $T_H$ be a transversal of $H$ in $\G$  and let $T_{K\le H}$ be a transversal of $K$ in $H$. Any element $g$ of $\G$ can be represented by $g=[g]_K+[g]_{T_{K\le H}}+[g]_{T_H}$ for unique $[g]_K\in K$, $[g]_{T_{K\le H}}\in T_{K\le H}$ and $[g]_{T_H}\in T_H$. Also note that $T_{K\le H}+T_H$ is a transversal $T_K$ of $K$ in $\G$ so that $g$ can be uniquely represented by $g=[g]_K+[g]_{T_K}$ for some $[g]_{T_K}\in T_K$ and $[g]_{T_K}$ can be uniquely represented by $[g]_{T_K}= [g]_{T_{K\le H}}+[g]_{T_H}$.

Given a source sequence $x_1^N\in\mathcal{X}^N$, the encoding rule is as follows: For $i\in[1,N]$, if $i\in A_{H,K}$ for some $K\le H\le \G$, $[v_i]_K$ is uniformly distributed over $K$ and is known to both the encoder and the decoder (and is independent from other random variables). The component $[v_i]_{T_{K\le H}}$ is the message and is uniformly distributed but is only known to the encoder. The component $[v_i]_{T_K}$ is chosen randomly so that for $g\in [v_i]_K+[v_i]_{T_{K\le H}}+T_H$,
\begin{align*}
P(v_i=g)=\frac{p_{V_i|X_1^NZ_1^NV_1^{i-1}}(g|x_1^N,z_1^N,v_1^{i-1})}{p_{V_i|X_1^NZ_1^NV_1^{i-1}}([v_i]_K+[v_i]_{T_{K\le H}}+T_H|x_1^N,z_1^N,v_1^{i-1})}
\end{align*}
For $i\in[1,N]$, if $i\in A_{H,K}$ for some $K\nleq H$, $[v_i]_H$ is uniformly distributed over $H$ and is known to both the encoder and the decoder and the component $[v_i]_{T_H}$ is chosen randomly so that for $g\in [v_i]_H+T_H$,
\begin{align*}
P(v_i=g)=\frac{p_{V_i|X_1^NZ_1^NV_1^{i-1}}(g|x_1^N,z_1^N,v_1^{i-1})}{p_{V_i|X_1^NZ_1^NV_1^{i-1}}([v_i]_H+T_H|x_1^N,z_1^N,v_1^{i-1})}
\end{align*}
For the moment assume that in this case $v_i$ is known at the receiver. Note that for $i\in[1,N]$, if $i\in A_{H,K}$ for some $K\le H\le \G$, $v_i$ can be decomposed as $v_i=[v_i]_K+[v_i]_{T_{K\le H}}+[v_i]_{T_H}$ in which $[v_i]_K$ is known to the decoder. The decoding rule is as follows: Given $z_1^N$ and for $i\in A_{H,K}$ for some $K\le H\le \G$, let
\begin{align*}
\hat{v}_i=\argmax_{g\in [v_i]_K+[v_i]_{T_{K\le H}}+T_H} W_{c,N}^{(i)}(z_1^N,\hat{v}_1^{i-1}|g)
\end{align*}
It is shown in the next section that with this encoding and decoding rules, the probability of error goes to zero. It remains to send the $v_i$ $i\in A_{H,K}$ with $K\nleq H$ to the decoder which can be done using a regular polar code (which achieves the symmetric capacity of the channel). Note that since the fraction $\frac{|A_{H,K}|}{N}$ vanishes as $N$ increases if $K\nleq H$, the rate loss due to this transmission can be made arbitrarily small.

\bibliographystyle{IEEEtran}
\bibliography{IEEEabrv,ariabib}
\end{document}